\documentclass{ws-procs961x669}            

\begin{document}

\title{ Lorentz invariance relations for  twist-3 quark distributions}

\author{Fatma P. Aslan}
 \address{Department of Physics, University of Connecticut, Storrs, CT 06269, U.S.A.}

\author{Matthias Burkardt}
 \address{Department of Physics, New Mexico State University,
Las Cruces, New Mexico,  88003-0001, USA}
\date{\today}

\begin{abstract}
We calculate twist-3 parton ditribution functions (PDFs) using cut and uncut diagrams. Uncut diagrams lead to a Dirac delta function term. No such term appears when cut diagrams are used. We show that a $\delta(x)$  is necessary to satisfy the Lorentz invariance relations of  twist-3 PDFs, except for the Burkhardt-Cottingham sum rule in QCD.
\end{abstract}


\bodymatter

\vspace*{-.1cm}
\section{Introduction} \label{s:intro}


In the scalar diquark model (SDM) and quark target model (QTM), twist-3 generalized parton distributions (GPDs) exhibit discontinuities at the points where the DGLAP and ERBL regions meet ($x =\pm\xi$)\cite {Aslan:2018zzk}.  In the forward limit, these discontinuities can grow into Dirac delta functions ($\delta(x)$) \cite{Aslan:2018zzk,Aslan:2018tff}. 
While none of the twist-2 PDFs exhibit these types of singularities in both models, all twist-3 PDFs, with the exception of $g_2(x)$ in the QTM, contain such singularities.
As we will show in section \ref{section:sumrules}, this $\delta(x)$  is necessary to satisfy the Lorentz invariance relations and the sum rules for twist-3 PDFs, except the Burkhardt-Cottingham sum rule in QCD \cite{Burkhardt:1970ti}.

This paper is organized as follows; in section \ref{section:LIR}, we investigate the two methods to calculate the PDFs, 'cut' and 'uncut' diagrams, and show that there is difference between the two approaches and one violates Lorentz invariance relations (LIR). Violations of sum rules involving higher twist PDFs is investigated in section \ref{section:sumrules}.


\vspace*{-.2cm}
\section{Lorentz invarince relations}\label{section:LIR}

Lorentz invariance, applied to the integral $I^\mu \equiv \int d^4k \hspace{.15cm}  \dfrac{k^{\mu}}{(k^2-m^2)^2}\delta[(P-k)^2- \lambda^2]$
implies $I^\mu \propto P^\mu$ as $P^\mu$ is the only 4-vector in this problem. Thus for $I^{\mu}$ the appropriate Lorentz invariance relation (LIR) reads,
\begin{equation}\label{LIR0}
\dfrac{I^+}{P^+}-\dfrac{I^-}{P^-}=0.
\end{equation}
\begin{figure}[!h]
	\begin{minipage}[t]{4cm}
		\includegraphics[scale=0.4]{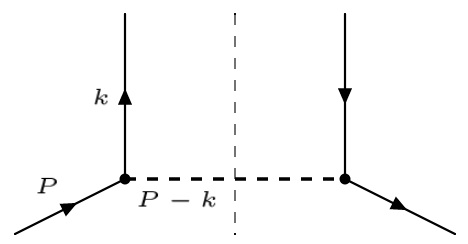}
		\caption{Cut diagram}
		\label{fig:cut}
	\end{minipage}
	\hspace{3cm}
	\begin{minipage}[t]{4cm}
		\includegraphics[scale=0.4]{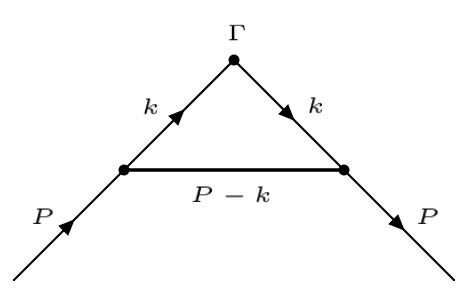}
		\caption{Uncut diagram}
		\label{fig:uncut}
	\end{minipage}
\end{figure}
 In the following, we will analyze this LIR in the SDM (in which the three valence quarks of the nucleon are considered to be in a bound state of a single quark and a scalar diquark)  using both
cut diagrams  and using uncut diagrams. We pretend that we are analysing a PDF where the factor
$k^\mu$ arises from the Dirac numerators.

In the
forward limit, the model can be represented using a cut diagram as in  FIG.\ref{fig:cut} or  an uncut diagram as in FIG. \ref{fig:uncut}.  
Using cut diagrams, the spectator propagator is replaced by $\delta\left( (p-k)^2-\lambda^2\right)$ thus enforcing the mass-shell
condition. Using uncut diagrams for the spectator line, the usual Feynman propagator is used and the energy integrals are performed using complex contour integration - picking up the pole of the spectator propagator.
Naively, the two methods should thus yield the same results. However, a subtle difference may arise at $x=0$ corresponding to infinite light-cone energy for the active quark.  In the literature, PDFs are calculated using either diagram. As we will show that, even though, both methods are equivalent and yield identical PDFs for nonzero $x$, a difference manifests itself at higher orders. This difference originates from the $\delta(x)$ term  which is present only in the higher order PDFs and  revealed when an uncut diagram is used. Such a term is not present in the calculations made by using cut diagrams! As we shall show, the $\delta(x)$ term is essential to satisfy the LIR involving twist-3 distributions and therefore the right approach is  to calculate higher order PDFs is to use uncut diagrams.

\vspace*{-.3cm}
\subsection{Cut Diagrams} \label{section:cut}

When cut diagrams are used,  $I^{\mu}\equiv I^{\mu}_{cut}$  is obtained as,
\begin{equation}\label{Imu}
I^{\mu}_{cut}\equiv P^+
\int_0^1 dx \hspace{.15cm} \int d^4k \hspace{.15cm}  \delta(k^+-xP^+) \dfrac{k^{\mu}}{(k^2-m^2)^2}\delta[(P-k)^2-\lambda^2].
\end{equation}
Here $P$  is the nucleon, $k$ is the quark momentum, $M, m, \lambda$ are the nucleon, quark and scalar diquark mass respectively.
%
The $k^+$ integral in Eq. (\ref{Imu}) is evaluated using $ \delta(k^+-xP^+) $,  while the $k^-$ intergral is evaluated by using the identity
\begin{equation}
\delta[(P-k)^2-\lambda^2]=\dfrac{1}{2P^+(1-x)}\delta\bigg(k^--\dfrac{M^2}{2P^+}+\dfrac{k_{\perp}^2+\lambda^2}{2P^+(1-x)}\bigg).
\end{equation}
Consequently, one finds for  $\mu=+$  and $\mu=-$ respectively,
\begin{equation}
\dfrac{I^+_{cut}}{P^+}\equiv
\dfrac{1}{2}\int{}d^2k_{\perp}\int_0^1{}dx \dfrac{x(1-x)}{(k_{\perp}^2+\omega)^2},
\end{equation}\vspace*{-.2cm}
\begin{equation}
\dfrac{I^-_{cut}}{P^-}\equiv
\dfrac{1}{P^-}\int{}\!\!\!d^2k_{\perp}\int_0^1{}\!\!\!dx\dfrac{M^2(1-x)-k_{\perp}^2-\lambda^2}{4P^+(k_{\perp}^2+\omega)^2},
\end{equation} where, 
$
\omega=-x(1-x)M^2+(1-x)m^2+x\lambda^2.
$
Hence, the LIR(\ref{LIR0}) is violated
\begin{eqnarray}\label{Difference}
\!\!\!\!\!\!\dfrac{I^+_{cut}}{P^+}-\dfrac{I^-_{cut}}{P^-}=\dfrac{1}{2M^2}\!\!\!\int{}\!\!d^2k_{\perp}\!\!\!\int_0^1{}\!\!\!\!dx \dfrac{-(1-x)^2M^2+k_{\perp}^2+\lambda^2}{(k_{\perp}^2+\omega)^2}=\dfrac{1}{2M^2}\!\!\int{}\!\!d^2k_{\perp}\dfrac{1}{k_{\perp}^2+m^2}.
\end{eqnarray}

\subsection{Uncut Diagrams}

However when  uncut diagrams are used, where $I^{\mu}\equiv I^{\mu}_{uncut}$ is defined as,
\begin{equation}\label{Imuncut}
I^{\mu}_{uncut}\equiv P^+
\int_0^1 dx \int d^4k \hspace{.15cm} \delta(k^+-xP^+) \dfrac{k^{\mu}}{(k^2-m^2+i\epsilon)^2}\dfrac{i}{[(P-k)^2-\lambda^2+i\epsilon]}.
\end{equation}


The $k^+$ integral  in (\ref{Imuncut}) is again taken using $ \delta(k^+-xP^+)$. However, in this case, the $k^-$ integral is taken using residue method.
For $\mu=+$, we obtain,
\begin{equation}\label{I+uncut}
\dfrac{I^+_{uncut}}{P^+}=\pi\int{}d^2k_{\perp}\int_0^1{}dx\dfrac{x(1-x)}{(k_{\perp}^2+\omega)^2}.
\end{equation}
For $\mu=-$, before taking the $k^-$ integral, we use the algebraic identity to rewrite the term in the numerator,
\begin{eqnarray}\label{kminus}
k^-= \dfrac{M^2}{2P^+}-\dfrac{(k_{\perp}^2+\lambda^2)}{2P^+(1-x)}-\dfrac{[(P-k)^2-\lambda^2]}{2P^+(1-x)}.
\end{eqnarray}
The last term in Eq.(\ref{kminus}) cancels the spectator propagator in the denominator leading to two different types of $k^-$ integrals in the expression for $I^-_{uncut}$,
\begin{equation}\label{Iminus2}
I^-_{uncut}\!
=\!\dfrac{i}{2P^+}\!\! \int_0^1{}\!\!\!\!\dfrac{dx}{1-x}\!\!\int{}\!\!d^2k_{\perp} \!\Big\{\int \!dk^-\!  \dfrac{M^2(1-x)-k_\perp^2-\lambda^2}{(k^2-m^2+i\epsilon)^2[(P\!-\!k)^2-\lambda^2+i\epsilon]}
- \int\! \!\dfrac{dk^-}{(k^2-m^2+i\epsilon)^2}\Big\}
\end{equation}
  The $k^-$ integral in Eq.(\ref{Iminus2}), leads to a delta function   \cite{Yan:1973qg},
\begin{eqnarray}
\int{}\dfrac{dk^-}{(k^2-m^2+i\epsilon)^2}=\dfrac{i\pi}{k_{\perp}^2+m^2}\delta(k^+).
\end{eqnarray}
Using this result, and taking the $k^-$ integrals in Eq.(\ref{Iminus2}) we obtain,
\begin{align}\label{I-uncut}
\dfrac{I^-_{uncut}}{P^+}=\dfrac{\pi}{2P^{+2}}\int{}d^2k_{\perp}\Big[\int_0^1{}dx\dfrac{M^2(1-x)-k_{\perp}^2-\lambda^2}{(k_{\perp}^2+\omega)^2}+\dfrac{1}{k_{\perp}^2+m^2}\Big],
\end{align}
which equals Eq. (\ref{I+uncut}), i.e. the
LIR (\ref{LIR0})  is satisfied when uncut diagrams are used.

The reason one method results in a violation of the  LIR while other does not
is the appearance of $\delta(x)$ term which is revealed only when an uncut diagram is used. Cut diagrams do not include the point $x=0$, and miss the $\delta(x)$ term at this point.


\section{Violation Of Sum Rules}\label{section:sumrules}

The point $x=0$ is not experimentally accessible in DIS  since  it corresponds to $ P \cdot q \rightarrow \infty$ and thus $\delta (x)$ cannot be seen. Any relation involving a twist 3 PDF containing a $\delta(x)$  would appear to be violated. Nevertheless, there  is no doubt in the validity of these sum rules because they are direct consequences of Lorentz invariance.  Therefore, the violation of the sum rules from the experimental data would provide an indirect evidence on the existence of the Dirac delta functions.

 The most famous Lorentz invariance relation between a twist 2 PDF ($g_1(x)$) and a twist 3 PDF ($g_T(x)$) is the
 Burkhardt-Cottingham sum rule \cite{Burkhardt:1970ti},
\begin{eqnarray}\label{BCsumrule}
\int_{-1}^1{}dx g_1 (x)=\int_{-1}^1{}dx g_T (x).
\end{eqnarray}
A similar relation is the $h$-sum rule, where the l.h.s. is equal to the tensor charge
\begin{eqnarray}\label{hsumrule}
\int_{-1}^1{}dx h_1 (x)=\int_{-1}^1{}dx h_L (x).
\end{eqnarray}

Another sum rule including a twist-3 PDF is the $\sigma$-term sum rule,  which provides a relation between quark mass $m$ and nucleon mass $M$,

\begin{eqnarray}\label{esumrule}
\int_{-1}^1{}dx e (x)=\dfrac{1}{2M}\langle P| \overline{\psi}(0)\psi(0)|P\rangle=\dfrac{d}{dm}M.
\end{eqnarray}
If any of the twist 3 PDFs above contain a $\delta (x)$ term, experimental measurements would not be able to confirm the sum rule in ) and claim their violation.


\section{Summary and Discussion}\label{section:sum}



Twist-3 PDFs contain a $\delta(x)$ in both QTM and SDM  only with the exception of $g_2 (x)$ in the QTM. These $\delta(x)$ terms are not related to the twist-2 (WW) parts of the twist-3 PDFs but contributes both the qgq correlation and mass terms. Since $x=0$ is not experimentally accesible, violations of the sum rules containing twist-3 PDFs and GPDs from the experimental data would provide an indirect evidence on the existence of these $\delta(x)$ contributions.

Using cut or uncut diagrams to calculate PDFs from models leads to the same result for $x\neq 0$. However, there is a difference between the two approaches for higher orders. Cut diagrams exclude $x=0$ point and hence miss the $\delta(x)$ terms. Therefore higher twist distributions calculated with cut digrams do not satisfy LIR. In order to restore it, one needs to include $x=0$ point by using an uncut diagram.

{\bf Acknowledgments}
This work was partially supported by the DOE under  Grant  No.DE-FG03-95ER40965.   F.A. was partially also supported by the DOE Contract No. DE-AC05-06OR23177, under which
Jefferson Science Associates, LLC operates Jefferson Lab.





\end{document}